# A critical review of the proposed reforms of the academic performance indicators applied in the assessment of researchers' performance in Hungary

**György Csomós**
University of Debrecen
4028 Debrecen, Hungary

**Abstract**
The academic performance indicators of the Doctor of Science title, the highest and most prestigious qualification awarded by the Hungarian Academy of Sciences (HAS), are key in the national assessment system. The types of performance indicators, as well as their minimum values, are incorporated into the application requirements of academic promotions, scientific qualifications, and research scholarships. HAS has proposed a reform of these performance indicators, to align with the current national and global trends. The proposed modifications are generally based on arbitrary decisions and the consensus between academicians, namely, the representatives of the sections of HAS. This paper contains a bibliometric analysis of 25,000 publications produced between 2011 and 2020 by 683 researchers affiliated with HAS's Section of Earth Sciences. The bibliometric data of the publications are processed by integer and fractional counting, respectively. The main goal of the paper is to argue that discipline-specific co-authorship patterns should be accounted for in the assessment procedure. It is also shown that the homogenization of the performance indicators and the rigid use of the integer counting method favour hard natural science disciplines and disadvantage social science disciplines. Finally, I describe some components of an alternative publishing strategy which would be most prudent for researchers, given the proposed assessment criteria.

## 1. Introduction

The assessment of individual researchers' performance is key when they apply for research grants, scientific qualifications, and academic promotion. Assessment practices vary from country to country, and from research field to research field, but it is generally accepted that scientometric indicators are the cornerstones of performance assessment (Abramo et al. 2013; Bloch and Schneider 2016; De Rijcke et al. 2016; Bornmann 2017). Because publications are considered to be one of the most common outcomes of research activity, performance assessments primarily focus on the quantity, the quality and the impact of publications (Sahel 2011; Bornmann and Marx 2014). Naturally, multiple indicators are available to evaluate individual researchers' performance, but in general, the number of published papers, the number of citations and the h-index are considered the fundamental indicators of one's scientific achievements (Coomes et al. 2013; Vavryčuk 2018). In some cases, additional indicators are added, such as the cumulative impact factor (Brito and Rodríguez-Navarro 2019; McKiernan et al. 2019; Zhang et al. 2017) and the number of articles published in the most prestigious journals in a particular field.

In Hungary, the assessment of individual researchers' performance is extremely metrics-based (as it has always been), and combines multiple academic performance indicators, measuring the quantity, quality and impact of publications. In general, the assessment procedure employs a wide range of performance indicators when a researcher applies for promotion (e.g., to obtain university professorship), research grants (e.g., to gain funding from OTKA, the largest basic research funding program in Hungary), scholarships (e.g., the János Bolyai research scholarship available for young researchers) or a higher academic qualification (e.g., the Habilitation and Doctor of Science titles) (Csomós, 2020).

Depending on the goal of the applications, different organizations handle and supervise the assessment procedure. The academic promotion procedures are handled by universities, except for the assessment of applications for the university professor position, which is carried out by the Hungarian Accreditation Committee (HAC), a national-level, independent body of experts tasked with the external evaluation of such applications. The OTKA basic research funding program is coordinated and supervised by the National Research, Development and Innovation Office, a governmental organization operating under the umbrella of the Ministry of Innovation and Technology. The János Bolyai research scholarship is the most important and prestigious scholarship established for researchers under the age of 45 by the Hungarian Academy of Sciences (HAS). The PhD degree and the Habilitation title are awarded by universities according to their in-house requirements, whereas the Doctor of Science (DSc) title, the highest and most prestigious scientific qualification in Hungary, is awarded by HAS.

At first, it is hard to recognize any commonalities in the assessment criteria of these different organisations. But if we dig deeper, it turns out that each of them is impacted by the academic performance indicators of the DSc title. The academic criteria of university professorship applications are directly based upon the field-specific performance indicators of the DSc title; to obtain the Habilitation title, a particular percentage of the performance indicators' minimum values must be achieved; the requirements of the János Bolyai research scholarship and the OTKA research grant loosely consider the performance indicators of the DSc title. In short, the academic performance indicators for the DSc title influence a broad range of important academic decisions in Hungary, regarding promotions, higher qualifications, research grants and scholarships.

To be in line with the global trends (e.g., the effort to remove the journal impact factor from the assessment criteria) and the changing publishing practices of researchers, the types and the minimum values of the performance indicators of the DSc title are renewed and recalibrated from time to time. However, even in the most radical revisions of these indicators, one sensitive topic is always neglected: using fractional counting in the assessment procedure to better reflect increases in co-authorship.

Of HAS's 11 sections, only the Section of Physical Sciences (which includes such disciplines as atomic and molecular physics, particle physics and nuclear physics) takes fractionally counted publication data into account to evaluate an applicant's research performance. This should come as no surprise, given that physics research tends to produce the publications with the highest number of co-authors (sometimes characterized as hyperauthorship) Cronin 2001; Castelvecchi 2015. It should be noted, however, that not even the Section of Physical Sciences employs the fractional counting method according to the 1/N rule (de Mesnard 2017; Hagen 2014; Osório 2018), but rather classifies the publications into arbitrary defined co-authorship categories (e.g., the number publications with 1-5 authors, 6-10 authors, …, more than 100 authors). The remaining 10 sections, including the Section of Medical Sciences, and some other sections containing natural science disciplines (e.g., the Section of Chemical Sciences and the Section of Biological Sciences), do not use the fractional counting approach at all.

The absence of an effective counting method may initially seem unproblematic, considering the fact that each section encompasses related disciplines. (For example, each of the 15 disciplines encompassed by the Section of Agricultural Sciences can be classified as an agricultural science discipline). The truth is, however, that most sections are characterized by significant discipline-specific differences in the average number of co-authors producing the scientific publications. This phenomenon results in a bias towards some disciplines in terms of productivity and citation impact, restricts the use of homogenized academic performance indicators on the section level, and can generate tension between researchers affiliated with a particular section. Naturally, the decision-makers of the sections (generally, the academicians) are aware of the fact that the co-authorship issue is an important factor in the establishment of the performance criteria. Yet, instead of conducting a straightforward bibliometric analysis to reveal the discipline-specific co-authorship pattern, the recalibration of the performance indicators is generally based on intuitions and the consensus between the representatives of the disciplines (Csomós, 2020). Eventually, in one way or another, the outcome of this procedure (i.e., the academic performance indicators defined by the sections of HAS) impacts the entire assessment practice.

In this paper, I conduct a bibliometric analysis to explore the discipline-specific co-authorship patterns of nine disciplines within the Section of Earth Sciences, including such hard natural science disciplines as geophysics and geochemistry, as well as social geography, a discipline that is generally included among the social sciences. In the analysis, I investigate the research performance of 683 researchers affiliated with the section by assessing approximately 25,000 publications produced in the period of 2011-2020, by both integer and fractional counting. The main goal of the paper is to demonstrate that the discipline-specific co-authorship practice results in a significant bias favouring the hard natural science disciplines. I also argue that it is rather problematic to establish a homogenized academic performance indicator scheme that similarly applies for all the disciplines belonging to the section. Finally, I consider some possible impacts of the underestimated co-authorship issue, the less adequately chosen academic performance indicators, and the arbitrarily recalibrated minimum values on the publishing strategy of (young) researchers. Before doing this, I give some brief insights into the Section of Earth Sciences, and the Hungarian Scientific Bibliography, the main data source of the analysis and the counting methods that are generally used in the assessment of individual researchers' performance.

## 2. Data collection and methods

### 2.1. An insight into the Section of Earth Sciences

HAS contains 11 sections encompassing most of the disciplines of such broader fields as medical, natural, and social sciences, engineering and arts and humanities. The Section of Earth Sciences classifies researchers into nine main scientific committees each representing a particular discipline. According to HAS's public database disclosing researchers' personal information, as of March 19, 2021, a total of 887 individuals were affiliated with the Section of Earth Sciences. The analysis is based on the publication data of 683 researchers (out of the 887 affiliated members), each of whom has a publicly available profile in HSB and produced at least one publication in the period of 2011-2020 (see the summary statistics of the committees in Table 1).

Table 1. Summary statistics of the committees of the Section of Earth Sciences

| Committee | Number of researchers | Number of researchers with HSB profiles | Percentage of researchers with HSB profiles in the dataset |
|---|---|---|---|
| Geochemistry | 120 | 95 | 13.90 |
| Geodesy | 57 | 45 | 6.59 |
| Geology | 78 | 51 | 7.47 |
| Geophysics | 91 | 65 | 9.52 |
| Meteorology | 93 | 69 | 10.10 |
| Mining | 65 | 40 | 5.86 |
| Palaeontology | 51 | 37 | 5.42 |
| Physical Geography | 136 | 110 | 16.10 |
| Social Geography | 196 | 171 | 25.04 |

### 2.2. Hungarian Scientific Bibliography, the source of the publication data processed in the analysis

Hungarian Scientific Bibliography (HSB), developed by HAS, has been operating since 2009. The main goal of the creation of a national publication and citation database was to provide a platform for Hungarian researchers that could store and make publicly available the bibliographic data of any type of publication written by Hungarian researchers in any language. The HSB soon became the official source of bibliometric data of Hungarian researchers, and when a researcher applies for promotion or qualification, the official HSB report on the researcher's publications must be presented.

The main advantage of HSB is that it is not biased towards any particular language or document type, in contrast to Web of Science (WoS) and Scopus, both of which are demonstrably biased towards

English language journal articles (Mongeon and Paul-Hus 2016). For example, in the period of 2011-2020, only 20 percent of the publications produced by researchers affiliated with the Section of Earth Sciences were indexed in the WoS, with this percentage dropping to lower than 7 percent for social geographers' publications.

It must be added, however, that despite the recent upgrades of HSB, it still provides less than optimal conditions for bibliometric analyses than WoS and Scopus (i.e., considering the analytical tools offered by the databases, HSB seems more like a mere inventory). One major problem of HSB is that it does not contain a search tool for citation data that would allow one to produce an author-specific citation report; instead, the citation data for each researcher must be collected manually.

Finally, it is also problematic that the creation and updating of personal HSB profiles are voluntary. As a result, 204 members of the Section of Earth Sciences, most of whom are 'senior' researchers, do not have publicly available HSB profiles or have not uploaded records in their existing profile. In addition, it happens sometimes that the voluntarily deposited publication records contain inaccurate bibliographic data, or that some data are missing (for example, the record is equipped with the WoS link, but the Scopus link is missing). The problems regarding the bibliographic data should, in principle, be corrected by the local HSB administrator (generally, the local librarian). Unfortunately, he/she is not always aware of all the field-specific bibliographic data types and does not always have the time to investigate and collect the missing citation data.

Despite the problems of HSB, it is the only database that provides comprehensive bibliographic data on the publications produced by Hungarian researchers, and the citations those publications received.

The 683 researchers having publicly available HSB profiles, each with at least one uploaded publication record in the profile, produced a total number of 25,021 scientific publications in the period of 2011-2020.

## 2.3. The academic performance indicators of the DSc title in the Section of Earth Sciences

Each section of HAS establishes a collection of academic performance indicators which are to be applied during the assessment of applicants for the DSc title. Because the publishing practices vary from field to field, the sections are allowed to employ field-specific academic performance indicators, which vary in terms of types and minimum values. In some sections, tailored sets of indicators are introduced for the disciplines (or group of disciplines) being encompassed that may more optimally reflect the differences in the publishing practices of the disciplines (or groups of disciplines). In the case of the Section of Earth Sciences, three sets of indicators are currently available for the disciplines (Table 2). The section is now about to harmonize the types of the indicators and recalibrate their minimum values. If that proposal is implemented, the same requirements must be considered for every researcher affiliated with the section who applies for the DSc title (see Table 2). The main goals of the modification are to reduce the heterogeneity of the indicators, to set the minimum values in accord with current publishing trends and to remove the impact factor from among the indicators.

So, how does the assessment procedure work in practice? A researcher may only apply for the DSc title if he/she meets or exceeds the minimum value of each performance indicator. Naturally, applicants must also demonstrate the proper teaching and public activity, but adequacy with respect to the performance indicators constitutes a fundamental criterion. That is, if an applicant does not fulfil one of the indicators, the application will be rejected. Each applicant receives one point for achieving the minimum value of a particular indicator. It is required that some indicators or all of them be exceeded by a magnitude of two, three or more. Actually, nobody knows (even those evaluating the applications), how many times the minimum value of a particular indicator should be overfulfilled, making the entire assessment procedure slightly vague for the applicants.

Table 2. The current and proposed types and minimum values of the academic performance indicators employed by the Section of Earth Sciences

| | Current minimum values | | | Proposed minimum values** |
|---|---|---|---|---|
| | For the disciplines of geochemistry, mineralogy, petrology, geology, geophysics, meteorology and palaeontology | For the disciplines of mining, geodesy, geoinformatics and physical geography | For the discipline of social geography | For every discipline |
| Number of scientific publications | 30 | 30 | 40 | 40 |
| Number of scientific publications with first author position | 15 | 15 | 20 | 20 |
| Number of scientific publications since obtaining last scientific degree | 15 | 15 | 30 | - |
| Number of scientific books and monographies | - | - | 2 | - |
| Number of scientific publications published in a foreign language | - | - | 35 | - |
| Number of journal articles indexed in SCI/SSCI (WoS) and Scopus* | 12 | 8 | 6 | 15 |
| Number of journal articles indexed in SCI/SSCI (WoS) and Scopus since obtaining last scientific degree* | 6 | 4 | 3 | - |
| Number of independent citations | 150 | 120 | 150 | 180 |
| Number of independent citations located in SCI/SSCI (WoS) and Scopus* | 50 | 30 | - | 80 |
| Cumulative impact factor value | 8 | 4 | 2 | - |
| Hirsch index | 9 | 8 | 8 | 10 |

*According to the description of the performance indicators, during the assessment procedure, only those publications and citations can be considered that are listed in the Science Citation Index (SCI) and the Social Sciences Citation Index (SSCI) databases. In reality, however, the publication and citation data of the journal articles listed in the Arts & Humanities Citation Index and the Emerging Sources Citation Index are also considered.

**The minimum values of the performance indicators might slightly change until their official approval, but it does not affect the basic principles of the reform.

It seems to be a crucial development that the section does not intend to replace the 'cumulative impact factor value' performance indicator with another one that somehow reflects on the quality of the journals containing the applicants' papers. According to some unofficial information circulating among the members, the decision-makers investigated the possible incorporation of the Scimago Journal Rank (i.e., the quartile rank of journals) in the assessment procedure, but the idea has finally been dropped. Now, none of the performance indicators represents the quality of the journals containing the publications, and this fact could very well impact researchers' publishing strategy (especially in the case of young researchers).

## 2.4. The most frequently used counting methods in the performance assessment

When a researcher applies for a research grant, scientific qualification, or academic promotion, he/she is required to submit his/her full publication record and the summary scientometric indicators of his/her publications. As was mentioned earlier, in Hungary, the assessment of individual researchers' performance is based on the integer counting approach; co-authorship metrics are not considered. This means, for example, that a researcher who is one of the authors of a publication co-produced by a research team of 10 people is awarded one full credit for that publication, just like the author who produced a publication without involving co-workers. This procedure also applies for the other performance indicators, including the number of citations and the cumulative impact factor value.

The question is how fair this procedure is. According to Katz and Martin (1997) high productivity in terms of publication output correlates with high levels of collaboration. In addition, many studies have demonstrated that the publications produced by more authors generally attract more citations (Biscaro and Giupponi 2014; Bosquet and Combes 2013; Tahamtan et al. 2016; Vieira and Gomes 2010). These findings let us conclude that the researcher who frequently collaborates with many other researchers most probably produces more publications, and those publications receive more citations. It is also well-documented that despite the increasing number of co-authors in publications produced in social sciences, high collaboration primarily characterizes the natural and life science disciplines (Henriksen 2016). Thus, it is of key importance to incorporate the most optimal co-authorship credit allocation into the performance assessment (see, for example, Herz et al. 2020).

As a matter of fact, the distribution of the publication credits among co-authors has been extensively discussed by scholars, and many alternative methods have been offered to replace the integer counting method. Alternatives to the integer (full/total/whole) counting approach include single- (first/last/corresponding) author counting (Cole and Cole 1974, Kosmulski 2012; Lange 2001; Zhang 2009, arithmetic counting (Trenchard 1992; Trueba and Guerrero 2004; Van Hooydonk 1997), geometric counting (Egghe et al. 2000), harmonic counting (Hagen 2010) and axiomatic counting Stallings et al. 2013). However, the fractional counting method is believed to be the most widespread alternative counting method used in performance assessment (Gauffriau 2017; Lindsey 1980; Price 1981; Rousseau 1992; Van Hooydonk 1997). By employing the fractional counting method, the publication credits can be distributed in equal fractions to each co-author according to the following equation (Osório 2018): $c_i^n = \frac{1}{n}$, for all $i \in N$.

In their studies, Bouyssou and Marchant (2016), Gauffriau et al. (2008), Mutz and Daniel (2019), Perianes-Rodriguez et al. (2016) and Zhou and Leydesdorff (2011) also demonstrate the advantages of the fractional counting method in the research assessment.

According to Vavryčuk (2018), some scholars criticize the fractional counting approach, asserting that it discourages collaborations and devalues the significance of some major authors' positions (e.g., first/last/corresponding author) in co-authored publications. However, as Waltman and van Eck (2015) point out, none of these arguments is straightforwardly justified. All things considered, fractional counting is superior to integer counting, especially when one approach is needed to evaluate the performance of researchers who are engaged in different fields (e.g., social and natural sciences).

**2.5. The methods employed in this analysis**

The main goals of this analysis are to evaluate the performance of 683 researchers affiliated with the Section of Earth Sciences to demonstrate how poorly suited the integer counting approach is, and why it is more effective to employ the fractional counting method in the assessment of individual researchers' performance. In the analysis, 25,021 publications, produced by the section's members between 2011 and 2020, are collected and examined at both the individual and the committee level. For each publication, four indicators are investigated: the number of scientific publications, the number of journal articles indexed in WoS, the number of independent citations and the number of citations located in WoS.

It must be noted that, in contrast to this analysis, the Section of Earth Sciences also considers journal articles and citations located in Scopus. However, while this study's data collection was carried out, HSB's limitations made it impossible to find out whether a given citation was contained by only Scopus, or WoS, or both; hence, it was impossible to control for overlaps between WoS and Scopus data. Moreover, as Gavel and Iselid (2008) have shown, there is indeed a significant overlap between the contents of Scopus and WoS; for example, in 2016, 84 percent of the active titles in WoS were also indexed in Scopus. In addition, in 2015, the WoS launched the Emerging Sources Citation Index (ESCI), a new index to include peer-reviewed publications of regional importance and in emerging scientific fields. Due this development, the overlap in terms of publication and citation data of Scopus and WoS has further increased. That is, the consideration of only WoS data instead of WoS and Scopus data does not impact the results of the analysis.

In the case of each publication, the values of the four indicators are calculated by integer and fractional counting, respectively. When using fractional counting each value of a given publication is divided by the number of co-authors (e.g., a publication produced by 10 co-authors awards 0.1 points to each researcher). Then, the values per indicator are aggregated at the individual level. After calculating the total value of a particular indicator for a researcher, that value is compared with the required minimum value for that performance indicator (see the minimum values in Table 2). First, an example of the use of the integer counting approach is demonstrated: A researcher who produced 80 publications between 2011 and 2020, regardless of the number of the co-authors of the publications, would receive 2 points, because the minimum value of the performance indicator the 'number of scientific publications' must be 40 items. If he/she produced 60 publications in that period, then he/she would receive 1.5 points. However, when the fractional counting approach is employed, each of the 80 publications is divided by the number of the co-authors. That is, if we apply the 1/N rule for each of the 80 publications, the result can be 25.7 publications, for instance, for which the applicant receives 0.623 points.

After this, by summing up and averaging the values of the performance indicators, two cumulative performance point-values per researcher are produced, the first of which is calculated via integer counting and the second, via fractional counting.

Naturally, when someone applies for the DSc title, he/she should present the scientific achievements of his/her entire career. However, this analysis focuses on demonstrating that the current assessment system favours researchers being characterized by high co-authorship (which correlates with higher productivity and higher citation impact) and disadvantages researchers who generally work alone or with only a few co-authors. To achieve the main goal of this research, it seems sufficient to collect and analyse publication and citation data from a 10-year period and compare those data with the required minimum values.

Finally, based on the cumulative performance points, calculated by both integer counting and fractional counting, the researchers are ranked, and then classified into the categories of the top 25% researchers. The disciplinary composition of these groups based on the ratio of researchers affiliated with a particular committee (which, in turn, represents a particular discipline) is also presented. This analysis shows that the choice between integer and fractional counting significantly affects the disciplinary composition of the group of top researchers.

## 3. Results

### 3.1. Mapping the co-authorship pattern of Earth Science disciplines

The researchers affiliated with the Section of Earth Sciences produced 25,021 publications between 2011 and 2020. Of these, 18 percent (4,511) were single-author publications. The ratio of publications in terms of co-authorship is highest in the category of 3-5 co-authors: 38.88 percent of the publications produced by the section's members belong to this category. The second highest ratio of publications (19.62 percent) were produced by 6-10 co-authors, followed by single-authored publications and publications with two co-authors (having the ratios of 18.03 and 18.02 percent, respectively).

Table 3 shows that social geography has the highest ratio of single-authored publications (35.65 percent), in stark contrast to less than 6 percent in the case of geochemistry. In fact, 54 percent of the single-authored publications are produced by social geographers. In addition, almost 63 percent of the publications of social geographers are written by only a single author or two co-authors. In contrast, 53 percent of the publications of meteorologists have 3-5 co-authors. In geophysics, 11 percent of the publications are produced by 11-20 co-authors, which is higher than the ratio of single-authored publications (7.03 percent).

Table 3. The ratio of publications in terms of co-authorship by disciplines

|  | 1 author | 2 authors | 3-5 authors | 6-10 authors | 11-20 authors | 21-50 authors | 51-100 authors | 101-500 authors | over 501 authors |
|---|---|---|---|---|---|---|---|---|---|
| Geochemistry | 5.90 | 7.37 | 40.22 | 35.24 | 9.13 | 1.02 | 1.02 | 0.10 | 0.00 |
| Geodesy | 22.00 | 19.50 | 35.86 | 16.27 | 5.73 | 0.37 | 0.09 | 0.18 | 0.00 |
| Geology | 6.20 | 13.58 | 44.76 | 26.80 | 6.70 | 1.96 | 0.00 | 0.00 | 0.00 |
| Geophysics | 7.03 | 13.78 | 39.71 | 25.11 | 10.78 | 3.10 | 0.05 | 0.33 | 0.11 |
| Meteorology | 9.29 | 13.30 | 52.95 | 17.85 | 4.51 | 1.63 | 0.43 | 0.04 | 0.00 |
| Mining | 13.29 | 25.27 | 43.28 | 16.12 | 1.67 | 0.29 | 0.00 | 0.07 | 0.00 |
| Palaeontology | 15.71 | 16.64 | 41.36 | 19.05 | 6.23 | 0.84 | 0.19 | 0.00 | 0.00 |
| Physical Geography | 15.97 | 15.81 | 40.42 | 24.71 | 1.95 | 1.05 | 0.04 | 0.04 | 0.00 |
| Social Geography | 35.65 | 27.25 | 29.04 | 7.13 | 0.70 | 0.17 | 0.06 | 0.00 | 0.00 |

Only 1.47 percent of all publications produced by the section's members contain more than 20 co-authors, and there are only two publications with more than 500 co-authors (both of which belong to the field of geophysics).

We can conclude that such hard natural science disciplines as geochemistry, geology, geophysics, and meteorology are characterized by high co-authorship, whereas social geography, a discipline being internationally classified under social sciences, is characterized by low co-authorship and the dominance of single- and dual-author publications. This difference in the co-authorship pattern is one of the most statistically significant factors affecting the researchers' productivity and citation impact.

**3.2. The main features of publication and citation data by disciplines**

The most publications are produced by social geographers, which is not entirely surprising considering that 25 percent of the section's members are affiliated with the Committee on Social Geography (Table 1). The physical geographers are the second highest contributors in terms of the number of publications. 47 percent of all publications of the Section of Earth Sciences come from these two geography disciplines. However, the international visibility of the publications produced by social geographers is extremely low: less than 7 percent of their publications are accounted for in WoS (Table 4). This ratio is also low in the case of mining, and only slightly higher in the case of geodesy. In contrast, hard natural science disciplines are characterized by high international visibility; for example, almost 42 percent of the geochemistry publications are indexed in WoS. There is a hypothesis circulating among Hungarian geographers according to which the low international visibility of publications produced by social geographers is due to the discipline's relatively narrow research scope. In other words, it is believed that international journals are less interested in publishing papers that exclusively focus on Hungarian society and the Hungarian geographical space, because those papers will predictably produce low citation impact. It should be added, however, that this hypothesis has never been validated. In addition, it remains questionable why the ratio of WoS indexed publications is low in the case of mining (i.e., in the case of a discipline that is classified into natural sciences).

The high international visibility of publications produced by a particular discipline goes beyond co-authorship issues and is related to further factors such as differences in discipline-specific publishing practices, standards and requirements.

The ratio of uncited publications varies from discipline to discipline, but as Table 4 shows, it is a common feature of the disciplines that most publications were not cited. The lowest uncited publication rate characterizes such hard natural science disciplines as geochemistry and palaeontology. Interestingly, social geography also belongs to this category. The publications produced by social geographers receive the lowest average number of citations, which is approximately the half the per-publication citations enjoyed by publications in meteorology. The average number of citations per cited item is also relatively low in the case of the geodesy, mining, geophysics, and physical geography. If we observe the average number of citations per cited WoS publication by discipline, we see extremely low values for social geography and mining. The publications of social geographers receive an average of only one WoS citation, whereas palaeontology, meteorology and geochemistry all boast more than 10 WoS citations per publication.

Table 4. Summary statistics of the publications by disciplines

| Scientific Committee | Number of publications | Ratio of publications indexed in WoS | Ratio of uncited publications | Total number of citations | Average number of citations per cited item | Number of citations located in WoS | Average number of WoS citations per cited item |
|---|---|---|---|---|---|---|---|
| Geochemistry | 3,133 | 41.65 | 51.99 | 14,657 | 9.75 | 15,557 | 10.34 |
| Geodesy | 1,082 | 15.43 | 74.31 | 1,613 | 5.80 | 1,193 | 4.29 |
| Geology | 2,194 | 26.48 | 66.64 | 6,703 | 9.16 | 6,716 | 9.17 |
| Geophysics | 1,836 | 30.07 | 66.01 | 4,310 | 6.91 | 4,848 | 7.77 |
| Meteorology | 2,572 | 26.44 | 65.36 | 9,779 | 10.98 | 9,540 | 10.71 |
| Mining | 1,377 | 7.77 | 77.56 | 1,920 | 6.21 | 895 | 2.90 |
| Palaeontology | 1,076 | 41.26 | 50.56 | 5,074 | 9.54 | 6,089 | 11.45 |
| Physical Geography | 4,864 | 16.26 | 64.14 | 12,017 | 6.89 | 7,767 | 4.45 |
| Social Geography | 6,887 | 6.85 | 59.69 | 15,168 | 5.46 | 2,849 | 1.03 |

Considering the differences being experienced in the discipline-specific co-authorship pattern, the international visibility of publications and the citation impact of the publications, we can predict that the reform of the academic performance indicators (e.g., the homogenization of the minimum values) being proposed by the section will favour researchers engaged in hard natural science disciplines and will unduly disadvantage other researchers (the social geographers, in particular). To avoid this undesirable situation, the section should reconsider the incorporation of the fractional counting method into the assessment.

**3.3. Ranking and classifying researchers employing integer vs. fractional counting**

After processing the academic performance indicators (i.e., the number of scientific publications, the number of journal articles indexed in WoS, the number of independent citations and the number of citations located in WoS) of a given researcher, two types of cumulative performance points are calculated, one achieved by integer counting, the other by fractional counting. As an outcome of this procedure, two rankings are established. The main goal of creating these two types of ranking is to find out the differences in the disciplinary composition of the top researchers. First, the top researchers of the section should be defined. In this analysis, the top researchers are those who occupy the top 25% in the rankings. That is, in the case of each ranking, 171 researchers are considered. When computing the cumulative performance points by integer counting, the section's members achieve a total of 2055 points of which 66 percent are produced by the top-25% researchers, while the sum of the fractionally counted performance points is 584 of which 60 percent is produced by the top researchers.

First, the disciplinary composition of the top-25% researchers in terms of integer counting is presented. In this case, the geochemistry discipline has the most researchers (40), with a ratio of 23.39 percent. Based on the number of the researchers, geochemistry is followed by physical geography (29) and meteorology (24). The disciplines of geodesy and mining are represented by 3 and 4 researchers, respectively. With 22 individuals, the social geography represents 12.87 percent in the group of the top-25% researchers. We can conclude that the disciplines with low co-authorship (see Table 3) are underrepresented in the group of the top researchers.

We now turn to the disciplinary composition of the top-25% group based on the fractionally counted performance points. Here, with 58 researchers, the social geographers make 33.92 percent of the group, followed by the physical geographers with a ratio of 17.54 percent (Fig. 1). This means that the contribution of the two geography disciplines to the group of the top researchers exceeds 50 percent. The ratio of researchers engaged in geochemistry decreases to 12.28 percent (from 23.39 percent calculated by integer counting). The disciplines of geodesy and mining still produce the lowest ratio with 2.92 percent, respectively.

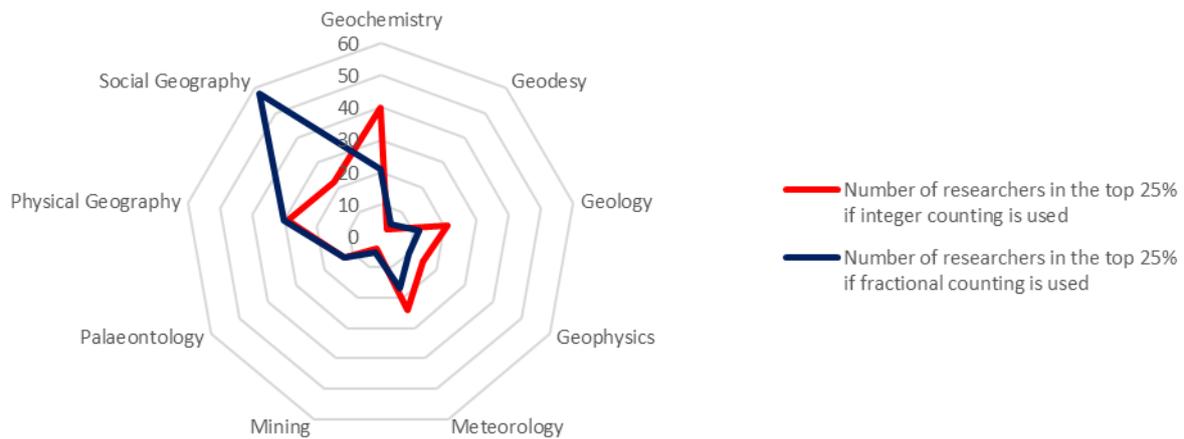

Fig. 1. The disciplinary composition of the top researchers affiliated with the Section of Earth Science using integer and fractional counting, respectively

When observing the ratios of researchers engaged in a particular discipline in the top-25% group, we see that the gap between the disciplines has narrowed when fractional counting was used instead of integer counting. As Fig. 2 shows, when using integer counting, both geology and geochemistry have more than 40 percent of their researchers counted in the group of the top-25% researchers. This ratio is the lowest for geodesy (6.67 percent) and mining (10.00 percent). In the case of social geography, only 12.87 percent of the discipline's researchers (22 researchers out of 171) will belong to the top-25% group.

The geodesy, mining and social geography disciplines are the main beneficiaries of the use of the fractional counting method, whereas some hard natural science disciplines being characterized by high co-authorship (e.g., geochemistry, geology and geophysics) experience decrease in the ratio of researchers belonging to the top group. For example, if we use integer counting, 40 out of 95 members of the Committee on Geochemistry will belong to the top-25% group but only 22 out of 171 members of the Committee on Social Geography will be included into this group. However, when fractional counting is used, the pattern will be significantly different: only 21 out of 95 members of the Committee on Geochemistry will be added to the top group, whereas the number of the members of the Committee on Social Geography will increase from 22 to 58.

Based on these findings we can assert that the section should take co-authorship into account to mitigate the bias towards hard natural science disciplines that currently characterizes the assessment procedure.

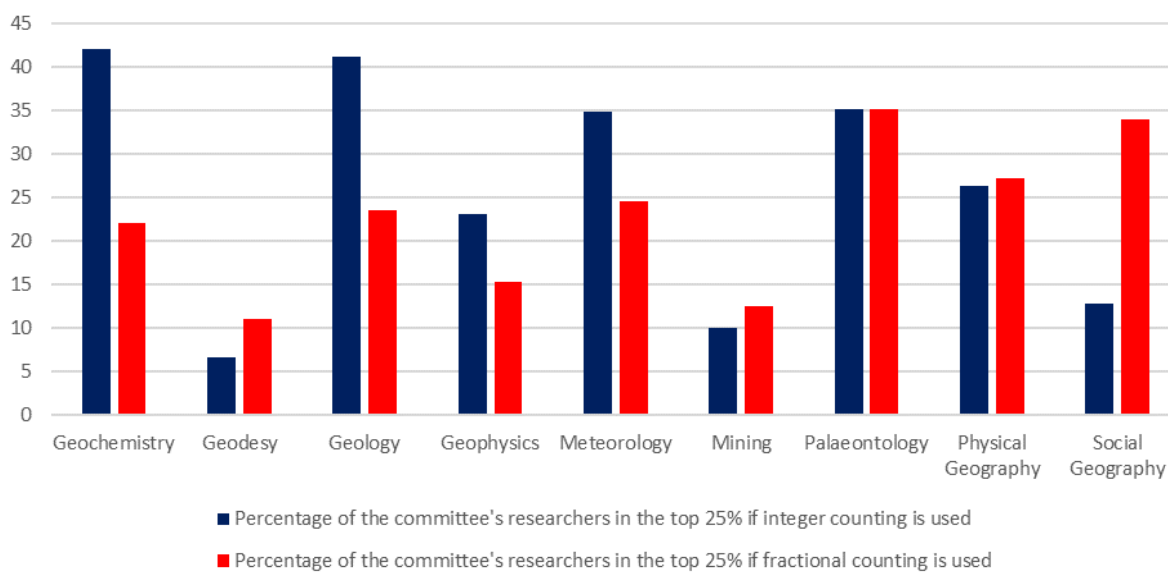

Fig. 2. The ratio of members of committees belonging to the top-25% researchers when using integer and fractional counting

## 4. Discussion and Conclusions

In Hungary, similarly to other Central and Eastern European post-socialist countries, the assessment of individual researchers' performance is highly metrics-based (Pajić 2015). The cornerstone of the entire assessment system is the Doctor of Science (DSc) title awarded by HAS. The DSc title is considered the most prestigious scientific qualification that a researcher may obtain during his/her career. Although, a faculty member is not required to hold the DSc title as a precondition of the application for such academic titles as the university professorship awarded by universities, the performance indicators of the DSc title are proportionally incorporated into the scientometric components of those applications. For example, if a faculty member applies for the university professor position, he/she is required to achieve the minimum values of the discipline-specific performance indicators of the DSc title. (A university professorship applicant is required to achieve 100 percent of the minimum values whereas for those applying for the Habilitation title, 60 percent of the minimum values must be achieved).

There are ongoing debates in the sections of HAS about the reform of the academic performance indicators. In the case of the Section of Earth Sciences, some of the proposed reforms, like the removal of the impact factor from among the indicators, aim to tailor the types of the performance indicators to the international trends (Cagan 2013; Hicks et al. 2015; Zhang et al. 2017), while others merely focus on the modification of the minimum values of the performance indicators. One of the main problems with the proposed modifications is that they are not based on a thorough and comprehensive bibliometric analysis but rather intuitions and arbitrary decisions. For example, the top-25% social geographers produce an average of 0.87 WoS publications per year, whereas the top physical geographers publish an average of 1.98 WoS items per year. One performance indicator, the number of journal articles indexed in WoS (and Scopus), is planned to require a minimum of 15 items for each discipline. It would take 17.2 years on average for the top social geographers to achieve that minimum value, in contrast to a mere 7.6 years for the top physical geographers. The minimum values of the performance indicators should be defined carefully because the disciplines can be characterized by significant differences in terms of publishing practice.

In addition, regardless of how progressive the performance indicators proposed by the section are, it seems that the consideration of the discipline-specific co-authorship pattern is the proverbial line that should never be crossed. This analysis demonstrates that the currently available performance indicator structure favours hard natural science disciplines against disciplines having a social science orientation. The incorporation of the fractional counting into the performance assessment seems to be highly reasonable as it would help narrow the gap between researchers being characterized by different discipline-specific productivity and citation impact.

If the homogenized performance indicator structure being proposed by the section (see Table 2) and the current evaluation methodology (i.e., the integer counting) were accepted and implemented, the publishing strategy of researchers (the young researchers in the first place) might change in response. In the following, some aspects of an alternative publishing strategy are presented.

1) Increase the team size to increase productivity and impact

A significant difference can be detected regarding the contribution of performance indicators to the cumulative performance point per discipline. In the case of geochemistry, meteorology and palaeontology, the main contributor to the cumulative performance point (over 40 percent) is the number of citations located in WoS. In contrast, for geodesy, mining and social geography, the number of scientific publications takes up the highest proportion of the cumulative performance point (with approximately 50 percent). The disciplines belonging to the latter group (social geography, in particular) lag behind in producing high impact.

Many studies have demonstrated that research collaborations help increase productivity and impact (He et al. 2009; Lee and Bozeman 2005). Natural science disciplines generally experience high collaboration (Ferligoj et al. 2015). Social sciences have also been witnessing an increasing tendency in co-authorship (Henriksen 2016; Ossenblok et al. 2014). This analysis demonstrates that social geography (as well as geodesy and mining) can be classified as one of the less collaborative disciplines. Social geographers tend to write publications on their own, or in cooperation with only one co-author, which might negatively affect the citation impact of the publications. That is, if social geographers team up and produce more publications being characterized by high co-authorship, they might be able to increase their competitiveness against researchers engaged in natural sciences. It may be ethically questionable to increase team size (see, for example, Bennett and Taylor 2003; Trinkle et al. 2017) even when this is not justified by the difficulty and complexity of the research. But it certainly is an effective way of increasing citation impact.

2) Neglect research topics that are more likely to produce low citation rates

As we compare the current minimum values of the performance indicators with the proposed ones (Table 2), we can see that each of them will experience increase. This development might generate a widening gap between researchers engaged in different disciplines. Due to the homogenization of the performance indicators, a meteorologist and a social geographer can satisfy the same assessment criteria even though their research generally has a very different impact in the scientific community. For example, a publication that focuses on climate change issues most probably attracts more citations and much more international citations than a publication dealing with Hungarian homesteads. The examination of this settlement type attracts a handful of social geographers whose publications can therefore collect only a few citations. As a result, it is more likely that a researcher who is committed to obtaining the DSc title, or just wants to climb higher on the career ladder, will not waste time and effort on investigating homesteads, but will instead focus on other research topics that have better chances of producing a higher citation rate.

That is, the proposed assessment criteria will lead to an uneven competition between researchers representing different disciplines, favour natural scientists and put social geographers into a disadvantageous position. The pressure on researchers to meet the extremely high assessment criteria may hijack them from some research topics that predictably produce low citation rates, regardless of how important those research topics are for society. (This issue is also investigated by Zhou et al. (2009) in the context of social vs. natural sciences).

3) Choose the easier way to save time and effort

The lack of indicators being established to demonstrate journal quality might motivate some researchers to avoid the submission of papers to highly ranked journals that are generally characterized by rigid reviewing process. Among the proposed indicators, the number of journal articles indexed in

WoS (and Scopus) is the only one that somehow reflects on the quality of journals containing the researchers' publications.

Scopus covers approximately 34,000 journals, and since the launching of the Emerging Sources Citation Index (ESCI), the coverage of WoS has increased to 21,000 journals. Now both Scopus and WoS contain some Hungarian earth science (and geography) journals that provide publishing spaces for Hungarian researchers in Hungarian as well as English. None of these journals is classified in the Q1 category as defined by the Scimago journal rank, and, except for *Acta Geodaetica et Geophysica*, a Q4 journal in terms of journal impact factor, all of them are listed by ESCI. That is, a researcher can easily bypass the stringent reviewing criteria imposed by the top journals if he/she publishes papers in such WoS or Scopus journals that are characterized by low rejection rate. The proposed assessment criteria draw no distinction between a paper published in Nature (or Science) and a Hungarian language paper published in a low impact WoS- or Scopus-indexed Hungarian journal. In short, the omission of performance indicators reflecting journal quality is very likely to be counterproductive and might negatively affect the international reputation of Hungarian earth scientists (at least in the case of some disciplines).

In conclusion, the types and the minimum values of the academic performance indicators of the DSc title awarded by HAS should be reconsidered, as should the counting method employed in the assessment of individual researchers' performance. The discipline-specific differences in productivity and impact should be considered in the assessment. Finally, and most importantly, the experts tasked with evaluating individual researchers' performance should look behind the curtain and examine the qualitative aspects of researchers' publications.